\documentstyle[12pt,qqa4lart]{article}
\input tcilatex

\QQQ{Language}{
American English
}

\begin{document}

\author{Lu-Ming Duan and Guang-Can Guo\thanks{%
Electronic address: gcguo@sunlx06.nsc.ustc.edu.cn} \\
Department of Physics and Nonlinear Science Center,\\
University of Science and Technology of China,\\
Hefei, Anhui 230026, People's Republic of China}
\title{Preserving coherence in quantum computation by pairing the quantum bits }
\date{}
\maketitle

\begin{abstract}
\baselineskip 24ptA scheme is proposed for protecting quantum states from
both independent decoherence and cooperative decoherence. The scheme
operates by pairing each qubit (two-state quantum system) with an ancilla
qubit and by encoding the states of the qubits into the corresponding
coherence-preserving states of the qubit-pairs. In this scheme, the
amplitude damping ( loss of energy) is prevented as well as the phase
damping (dephasing) by a strategy called the free-Hamiltonian-elimination We
further extend the scheme to include quantum gate operations and show that
loss and decoherence during the gate operations can also be prevented. \\

\end{abstract}

\newpage\baselineskip 24ptSoon after the idea of quantum computation became
an active part of current research through the innovative work of Shor on
factorization [1,2], decoherence was recognized as a major problem that can
not be ignored [3], especially when one is interested in practical
applications. Quantum computers act as sophisticated nonlinear
interferometers. The coherent interference pattern between the multitude of
superpositions is essential for taking advantage of quantum parallelism.
However, decoherence of the qubits caused by the interaction with
environment will collapse the state of the quantum computer and make the
information no longer correct. To overcome this fragility of quantum
information, Shor, and independently Steane, inspired by the theory of
classical error correction, proposed the first two quantum error-correcting
codes (QECCs), i.e., the 9-bit code [4] and the 7-bit code [5], which are
able to correct errors that occur during the storage of qubits. Furthermore,
a general theory for quantum error correction was presented by Calderbank
and Shor [6], and independently by Steane [7]. Following this work, many new
QECCs have since been discovered [8-21]. The discovery of QECCs has
revolutionized the field of quantum information.

Quantum errors are induced by the interaction of the qubits with
environment. If we know more about this interaction, simpler codes can be
found. In the previous analyses of decoherence [3], the qubits are assumed
to interact independently with separate environments. In practice, however,
cooperative effects may take place between the qubits. For example, the
qubits in the ion-trapped computers are believed to be decohered
cooperatively [22,23]. Refs. [24] and [25] considered another extreme case,
i.e., all the qubits interact with the same environment. If only the phase
damping is considered, as the result, the qubits are found to be decohered
collectively. For some of the input states (called the sub-decoherent
states), the qubits are decohered much slower; and for some others (called
the super-decoherent states), they are decohered much faster. The phenomenon
of super-decoherence vs. sub-decoherence is very similar to but not the same
as the process of super-radiance vs. sub-radiance more commonly encountered
in literature [26]. As was pointed out in Ref. [24], super-radiance is a
process of collective radiation by a group of closely spaced atoms, while
super-decoherence is due to collective entanglement between qubits and
environment. A simple code has been suggested in [24] for reducing this
collective decoherence.

Independent decoherence and collective decoherence are extreme cases. With
these two ideal circumstances, we ask , what about the real situation? It
seems a combination of these two cases may be more practical. If the qubits
are close, they tend to be decohered collectively; and if they are departed,
the assumption of independent decoherence may be more reasonable. In this
letter, we propose a scheme for reducing decoherence in general cases. The
scheme operates by pairing each qubit with an ancilla. The two qubits in
each pair are set close so that they interact with the same modes of the
environment. But the qubits in different pairs are allowed to be decohered
independently or cooperatively. Due to the collective dissipation in each
pair, coherence-preserving states of the qubit-pairs are found to exist. The
stored information is protected from decoherence by encoding the states of
the qubits into the corresponding coherence-preserving states of the
qubit-pairs. In fact, the use of coherence-preserving states for preventing
errors induced by the pure dephasing has been described by Chuang and
Yamamoto [27,28] and also by Palma, etc. [24]. Here we adopt the
previously-known idea of using such states of qubit-pairs. We propose a
strategy called the free-Hamiltonian-elimination to provide a general method
to set up the coherence-preserving states. By this strategy, the amplitude
damping is prevented as well as the phase damping. The amplitude damping
sometimes is a main source of decoherence [23,29,30]. Furthermore, we show
in this letter that the scheme can be extended to prevent decoherence in
quantum gate operations. Coherence is preserved in the gate operations by
substituting the logic gates for the qubits with those for the qubit-pairs.
Preserving coherence during quantum gate operations is a significant step
towards realizing the fault-tolerant quantum computation [15].

First, we consider the stored information, i.e., the qubits in quantum
memory, which can be described by Pauli's operators $\overrightarrow{\sigma }%
_l$ ( $l$ marks different qubits). The environment is modelled by a bath of
oscillators with infinite degrees of freedom. Each qubit interacts with some
( usually infinite) modes of the environment. The bath modes coupling with
the $l$ qubit are indicated by $a_{\omega l}$ ( $\omega $ varies from $0$ to 
$\infty $ ). For different $l_1$ and $l_2$, some of the modes $a_{\omega
l_1} $ and $a_{\omega l_2}$ are possibly the same and some of them are
different. We use the notation $\stackrel{L}{\stackunder{l=1}{\bigcup }}A_l$
to indicate the joint sum of $A_l$, where all $A_l$ are bath operators. For
example, $\stackrel{2}{\stackunder{l=1}{\bigcup }}A_l=A_1+A_2$ if $A_1$ and $%
A_2$ belong to different modes; and $\stackrel{2}{\stackunder{l=1}{\bigcup }}%
A_l=A_1$ if $A_1$ and $A_2$ are the same. With this notation, the whole
Hamiltonian describing the general dissipation of the qubits, including the
phase damping and the amplitude damping, has the following form ( setting $%
\hbar =1$ ) 
\begin{equation}
\label{1}
\begin{array}{c}
H_L=\omega _0 
\stackrel{L}{\stackunder{l=1}{\sum }}\sigma _l^z+\stackunder{\omega }{\sum }%
\stackrel{L}{\stackunder{l=1}{\bigcup }}\left( \omega a_{\omega
l}^{+}a_{\omega l}\right) \\  \\ 
+\stackrel{L}{\stackunder{l=1}{\sum }}\stackunder{\omega }{\sum }\left[
\left( \lambda ^{\left( 1\right) }\sigma _l^x+\lambda ^{\left( 2\right)
}\sigma _l^y+\lambda ^{\left( 3\right) }\sigma _l^z\right) g_{\omega
l}\left( a_{\omega l}^{+}+a_{\omega l}\right) \right] , 
\end{array}
\end{equation}
where $L$ is the number of qubits and the coupling constants $g_{\omega l}$
may be dependent of $\omega $ and $l$. The ratio $\lambda ^{\left( 1\right)
}:\lambda ^{\left( 2\right) }:\lambda ^{\left( 3\right) }$ is determined by
the type of the dissipation. For example, if $\lambda ^{\left( 1\right)
}=\lambda ^{\left( 2\right) }=0$, it describes the phase damping; and if $%
\lambda ^{\left( 3\right) }=0$, it is the amplitude damping.

Now we pair each qubit with an ancilla. The ancilla of the $l$ qubit is
indicated by $l^{^{\prime }}$. The two qubits $l$ and $l^{^{\prime }}$ in
the pair are set close so that they interact with the same modes of the
environment. With this condition, the dissipation of the $L$ qubit-pairs is
described by the Hamiltonian 
\begin{equation}
\label{2}
\begin{array}{c}
H_{2L}=\omega _0 
\stackrel{L}{\stackunder{l=1}{\sum }}\left( \sigma _l^z+\sigma _{l^{^{\prime
}}}^z\right) +\stackunder{\omega }{\sum }\stackrel{L}{\stackunder{l=1}{%
\bigcup }}\left( \omega a_{\omega l}^{+}a_{\omega l}\right) \\  \\ 
+\stackrel{L}{\stackunder{l=1}{\sum }}\stackunder{\omega }{\sum }\left\{
\left[ \lambda ^{\left( 1\right) }\left( \sigma _l^x+\sigma _{l^{^{\prime
}}}^x\right) +\lambda ^{\left( 2\right) }\left( \sigma _l^y+\sigma
_{l^{^{\prime }}}^y\right) +\lambda ^{\left( 3\right) }\left( \sigma
_l^z+\sigma _{l^{^{\prime }}}^z\right) \right] g_{\omega l}\left( a_{\omega
l}^{+}+a_{\omega l}\right) \right\} , 
\end{array}
\end{equation}

The following step of our strategy is to eliminate the influence of the free
Hamiltonian $H_0=\omega _0\stackrel{L}{\stackunder{l=1}{\sum }}\left( \sigma
_l^z+\sigma _{l^{^{\prime }}}^z\right) $ of the qubits. To attain this goal,
we introduce a homogeneous classical driving electromagnetic field which
acts on all the qubit-pairs. The ancillary Hamiltonian describing the
driving process is 
\begin{equation}
\label{3}H_{drv}=\stackrel{L}{\stackunder{l=1}{\sum }}\left[ g\left( \sigma
_l^{+}+\sigma _{l^{^{\prime }}}^{+}\right) +g^{*}\left( \sigma _l^{-}+\sigma
_{l^{^{\prime }}}^{-}\right) \right] =\stackrel{L}{\stackunder{l=1}{\sum }}%
\left[ g_1\left( \sigma _l^x+\sigma _{l^{^{\prime }}}^x\right) +g_2\left(
\sigma _l^y+\sigma _{l^{^{\prime }}}^y\right) \right] , 
\end{equation}
By adjusting the intensity and the phase of the driving field, we can choose
the driving constants $g_1$ and $g_2$ to satisfy $g_1:g_2:\omega _0=\lambda
^{\left( 1\right) }:\lambda ^{\left( 2\right) }:\lambda ^{\left( 3\right) }$%
. Then the whole Hamiltonian is simplified to 
\begin{equation}
\label{4}
\begin{array}{c}
H=H_{2L}+H_{drv} \\  
\\ 
=\stackrel{L}{\stackunder{l=1}{\sum }}\left\{ \left( S_l+S_{l^{^{\prime
}}}\right) \left[ \frac{\omega _0}{\lambda ^{\left( 3\right) }}+\stackunder{%
\omega }{\sum }g_{\omega l}\left( a_{\omega l}^{+}+a_{\omega l}\right)
\right] \right\} +\stackunder{\omega }{\sum }\stackrel{L}{\stackunder{l=1}{%
\bigcup }}\left( \omega a_{\omega l}^{+}a_{\omega l}\right) , 
\end{array}
\end{equation}
where we have let $S_l=\lambda ^{\left( 1\right) }\sigma _l^x+\lambda
^{\left( 2\right) }\sigma _l^y+\lambda ^{\left( 3\right) }\sigma _l^z.$

Suppose the initial state of the qubit-pairs is a co-eigenstate of all the
operators $S_l+S_{l^{^{\prime }}}$, with the eigenvalues $m_l$,
respectively. The environment state is indicated by $\left| \Psi
_{env}\left( 0\right) \right\rangle $. Under the Hamiltonian (4), at time $t$
the state of the whole system evolves into 
\begin{equation}
\label{5}
\begin{array}{c}
\left| \Psi \left( t\right) \right\rangle =e^{-iHt}\left( \left| \Psi \left(
0\right) \right\rangle \otimes \left| \Psi _{env}\left( 0\right)
\right\rangle \right) \\  
\\ 
=\left| \Psi \left( 0\right) \right\rangle \otimes e^{-it\left\{ \stackrel{L%
}{\stackunder{l=1}{\sum }}m_l\left[ \frac{\omega _0}{\lambda ^{\left(
3\right) }}+\stackunder{\omega }{\sum }g_{\omega l}\left( a_{\omega
l}^{+}+a_{\omega l}\right) \right] +\stackunder{\omega }{\sum }\stackrel{L}{%
\stackunder{l=1}{\bigcup }}\left( \omega a_{\omega l}^{+}a_{\omega l}\right)
\right\} }\left| \Psi _{env}\left( 0\right) \right\rangle . 
\end{array}
\end{equation}
So in this case all the qubit-pairs undergo no decoherence, though they are
interacting with the environment. Because of this property, we call the
eigenstates of all the operators $S_l+S_{l^{^{\prime }}}$ the
coherence-preserving states.

We briefly discuss the coherence-preserving states. The Hermitian operator $%
S_l$ satisfies $tr\left( S_l\right) =0$, so its two eigenstates, without
loss of generality, can be indicated by $\left| \pm 1\right\rangle _l$, with
the eigenvalues $\pm a$, respectively. The computation basis states $\left|
\pm \right\rangle _l$ are eigenstates of the operator $\sigma _l^z$. The
states $\left| \pm 1\right\rangle _l$ may differ with $\left| \pm
\right\rangle _l$ by a single-qubit rotation operation $R_l\left( \theta
\right) $, i.e., $\left| \pm 1\right\rangle _l=R_l\left( \theta \right)
\left| \pm \right\rangle _l$, where $\theta $ depends on the type of the
dissipation. The coherence-preserving states can be easily constructed from
the states $\left| \pm 1\right\rangle _l$. The largest eigen-space of the
operator $S_l+S_{l^{^{\prime }}}$ is a $2$-dimensional space spanned by the
eigenstates $\left| +1,-1\right\rangle _l$ and $\left| -1,+1\right\rangle _l$%
, with the eigenvalue $m_l=0$. So there exists a one-to-one map form the $2$%
-dimensional space of a qubit onto the $2$-dimensional coherence-preserving
state space of a qubit-pair. The general input states of $L$ qubits can be
expressed as 
\begin{equation}
\label{6}\left| \Psi _L\right\rangle =\stackunder{\left\{ i_l\right\} }{\sum 
}c_{\left\{ i_l\right\} }\left| \left\{ i_l\right\} \right\rangle , 
\end{equation}
where $\left\{ i_l\right\} $ is abbreviation of the notation $i_1,i_2,\cdots
,i_L$ and $i_l=\pm 1,$ $l=1,2,\cdots ,L$. We encode the state (6) into the
following coherence-preserving state of $L$ qubit-pairs 
\begin{equation}
\label{7}\left| \Psi _{2L}\right\rangle _{coh}=\stackunder{\left\{
i_l\right\} }{\sum }c_{\left\{ i_l\right\} }\left| \left\{ i_l,-i_l\right\}
\right\rangle , 
\end{equation}
where $\left\{ i_l,-i_l\right\} $ indicates $i_1,-i_1,i_2,-i_2\cdots
,i_L,-i_L$. The encoding can be fulfilled by the quantum CNOT
(Controlled-NOT) operations $C_{ij}$, where the first subscript of $C_{ij}$
refers to the control bit and the second to the target. The ancillas are
prearranged in the state $\left| \Psi _{1^{^{\prime }}2^{^{\prime }}\cdots
L^{^{\prime }}}\right\rangle =\left| +1\right\rangle _{1^{^{\prime
}}}\otimes \left| +1\right\rangle _{2^{^{\prime }}}\otimes \cdots \otimes
\left| +1\right\rangle _{L^{^{\prime }}}$. Let the joint operation $%
C_{ij}^{^{\prime }}\left( \theta \right) =R_i\left( \theta \right) R_j\left(
\theta \right) C_{ij}R_i\left( -\theta \right) R_j\left( -\theta \right) $,
where $R_i\left( \theta \right) $ is the rotation operation acting on the $i$
qubit, we thus have 
\begin{equation}
\label{8}\left| \Psi _L\right\rangle \otimes \left| \Psi _{1^{^{\prime
}}2^{^{\prime }}\cdots L^{^{\prime }}}\right\rangle \stackrel{%
C_{11^{^{\prime }}}^{^{\prime }}\left( \theta \right) C_{22^{^{\prime
}}}^{^{\prime }}\left( \theta \right) \cdots C_{LL^{^{\prime }}}^{^{\prime
}}\left( \theta \right) }{\longrightarrow \longrightarrow }\left| \Psi
_{2L}\right\rangle _{coh}. 
\end{equation}
The decoding can be similarly realized by applying the operation $%
C_{11^{^{\prime }}}^{^{\prime }}\left( \theta \right) C_{22^{^{\prime
}}}^{^{\prime }}\left( \theta \right) \cdots C_{LL^{^{\prime }}}^{^{\prime
}}\left( \theta \right) $ again. The encoded states $\left| \Psi
_{2L}\right\rangle _{coh}$ undergo no decoherence in the memory.

By pairing the qubits, the number of qubits is expanded from $L$ to $2L$. So
the efficiency $\eta $ of this scheme is $\frac 12$. There is a possible way
to raise the efficiency. If $2m$ qubits are set close so that they all
interact with the same modes of the environment, the largest eigen-space of
the operator $S_1+S_2+\cdots +S_{2m}$ becomes a $\left( 
\begin{array}{c}
2m \\ 
m 
\end{array}
\right) $-dimensional state space, with the eigenvalue $m_l=0$. By encoding
the input states of $2mL$ qubits into the coherence-preserving states of the
qubit-clusters, each cluster consisting of $2m$ qubits, the maximum
efficiency $\eta _m$ attains 
\begin{equation}
\label{9}\eta _m=\frac L{2mL}\log _2\left( 
\begin{array}{c}
2m \\ 
m 
\end{array}
\right) \approx 1-\frac 1{4m}\log _2\left( \pi m\right) , 
\end{equation}
where the approximation is taken under the condition $m>>1$. So the
efficiency $\eta _m$ is near to $1$ if $m$ is large. Of course, with $m$
increasing, it becomes harder and harder to set all the $m$ qubits close so
that they are decohered collectively.

In the above, we have dealt with the qubits in the memory. Now we extend the
scheme to include quantum gate operations. In quantum error-correction
schemes, a significant step forward in this direction has recently been made
by the idea of fault-tolerant implementation of quantum logic gates [15-17].
Here we show our coherence-preserving scheme can, at least in principle,
prevent decoherence during the gate operations as well as during the storing
process. The Hamiltonian for the gate operation is indicated by $H_g$. The
initial state $\left| \Psi \left( 0\right) \right\rangle _{\left\{
m_l\right\} }$ of the qubit-pairs is a co-eigenstate of all the operators $%
S_l+S_{l^{^{\prime }}}$, with the eigenvalue $m_l$, respectively. If the
gate Hamiltonian $H_g$ satisfies the following condition 
\begin{equation}
\label{10}\left[ H_g,S_l+S_{l^{^{\prime }}}\right] =n_l,\text{ }l=1,2,\cdots
,L\text{ }, 
\end{equation}
where all $n_l$ are numbers, at time $t$ the whole system, including the
environment, will evolve into 
\begin{equation}
\label{11}
\begin{array}{c}
\left| \Psi \left( t\right) \right\rangle =e^{-iH_gt}\left| \Psi \left(
0\right) \right\rangle _{\left\{ m_l\right\} } \\  
\\ 
\otimes e^{-it\left\{ \stackrel{L}{\stackunder{l=1}{\sum }}\left( m_l-\frac
12n_l\right) \left[ \frac{\omega _0}{\lambda ^{\left( 3\right) }}+%
\stackunder{\omega }{\sum }g_{\omega l}\left( a_{\omega l}^{+}+a_{\omega
l}\right) \right] +\stackunder{\omega }{\sum }\stackrel{L}{\stackunder{l=1}{%
\bigcup }}\left( \omega a_{\omega l}^{+}a_{\omega l}\right) \right\} }\left|
\Psi _{env}\left( 0\right) \right\rangle . 
\end{array}
\end{equation}
Therefore, in this case no decoherence occurs during the gate operation. Eq.
(10) is also a necessary condition for preserving coherence during the gate
operation.

Now we show, with the constraint (10), any unitary transformations can still
be constructed. To demonstrate this, we only need to give a universal gate
operation satisfying Eq. (10). It has been proven that almost any $2$-bit
gates are universal [31,32]. In particular, the following is a universal
gate operation [33] 
\begin{equation}
\label{12}U_{l_1l_2}=\left| -1\right\rangle _{l_1\text{ }l_1}\left\langle
-1\right| I_{l_2}+\left| +1\right\rangle _{l_1\text{ }l_1}\left\langle
+1\right| V_{l_2}, 
\end{equation}
where $I_{l_2}$ is a $2\times 2$ unit matrix and the unitary matrix $V_{l_2}$
is given by 
\begin{equation}
\label{13}V_{l_2}\left( \alpha ,\theta ,\phi \right) =\left( 
\begin{array}{cc}
e^{i\alpha }\cos \left( \theta \right) & -ie^{i\left( \alpha -\phi \right)
}\sin \left( \theta \right) \\ 
-ie^{i\left( \alpha +\phi \right) }\sin \left( \theta \right) & e^{i\alpha
}\cos \left( \theta \right) 
\end{array}
\right) . 
\end{equation}
The parameters $\alpha ,\theta ,\phi $ are irrational multiples of $\pi $
and of each other. Now we consider the following gate operation for two
qubit-pairs $l_1l_1^{^{\prime }},l_2l_2^{^{\prime }}$%
\begin{equation}
\label{14}U_{l_1l_1^{^{\prime }}l_2l_2^{^{\prime }}}=\left|
-1,+1\right\rangle _{l_1l_1^{^{\prime }}\text{ }l_1l_1^{^{\prime
}}}\left\langle -1,+1\right| I_{l_2l_2^{^{\prime }}}+\left|
+1,-1\right\rangle _{l_1l_1^{^{\prime }}\text{ }l_1l_1^{^{\prime
}}}\left\langle +1,-1\right| V_{l_2l_2^{^{\prime }}}, 
\end{equation}
where $I_{l_2l_2^{^{\prime }}}$ is a $4\times 4$ unit matrix and $%
V_{l_2l_2^{^{\prime }}}$ becomes ( in the basis\\ $\left\{ \left|
-1,-1\right\rangle ,\text{ }\left| -1,+1\right\rangle ,\text{ }\left|
+1,-1\right\rangle ,\text{ }\left| +1,+1\right\rangle \right\} $ ) 
\begin{equation}
\label{15}V_{l_2l_2^{^{\prime }}}\left( \alpha ,\theta ,\phi \right) =\left( 
\begin{array}{cccc}
1 &  &  &  \\  
& e^{i\alpha }\cos \left( \theta \right) & -ie^{i\left( \alpha -\phi \right)
}\sin \left( \theta \right) &  \\  
& -ie^{i\left( \alpha +\phi \right) }\sin \left( \theta \right) & e^{i\alpha
}\cos \left( \theta \right) &  \\  
&  &  & 1 
\end{array}
\right) . 
\end{equation}
After decoding the coherence-preserving states of the qubit-pairs into the
original states of the qubits, the operation (14) for the qubit-pairs just
corresponds to the operation (12) for the qubits. So Eq.(14) gives a
universal gate operation for the qubit-pairs. For any parameters $\alpha
,\theta ,\phi $, it is easy to check that $U_{l_1l_1^{^{\prime
}}l_2l_2^{^{\prime }}}$ satisfies 
\begin{equation}
\label{16}\left[ U_{l_1l_1^{^{\prime }}l_2l_2^{^{\prime
}}},S_{l_1}+S_{l_1^{^{\prime }}}\right] =\left[ U_{l_1l_1^{^{\prime
}}l_2l_2^{^{\prime }}},S_{l_2}+S_{l_2^{^{\prime }}}\right] =0, 
\end{equation}
so the generators of $U_{l_1l_1^{^{\prime }}l_2l_2^{^{\prime }}}$ , i.e.,
the gate Hamiltonians, also commute with the operators $S_l+S_{l^{^{\prime
}}}$. The constraint (10) is therefore satisfied.

In the above, we have shown coherence can be preserved during gate
operations if one substitutes the gates for the qubits with those for the
qubit-pairs. Of course, after this substitution, the demonstration of these
logic gates becomes more involved.

Finally, we compare this scheme with quantum error correction. In the error
correction schemes, the decoherence time for a qubit is not increased. What
one does is to retrieve the useful information from the decohered state by
introducing some redundancy. Contrary to this, in our scheme, the
decoherence time for the qubits is much increased. ( In the ideal case, it
is increased to infinity. ) We prevent error rather than correct error. So,
like Ref. [34,35], this scheme belongs to the class of error prevention
schemes. The schemes of Ref. [34,35] are based on the quantum Zeno effect.
The decoherence is reduced by continuously measuring the qubits in some
basis. The critical idea of our scheme is pairing the qubits and
substituting the gate operations for the qubits with those for the
qubit-pairs. This scheme has some attractive features. First, it covers a
large range of decoherence., including the cooperative decoherence and the
independent decoherence. The scheme works whether the decoherence is caused
by the amplitude damping or by the phase damping. Second, it has a high
efficiency. We need at most two qubits to encode a qubit. Third, the
encoding and the decoding in this scheme is quite simple. It only needs $L$
times quantum CNOT operations and some single-bit rotation operations to
encode and decode the qubits. Last, the scheme is relatively easy to extend
for preventing decoherence in quantum gate operations. Of course, compared
with QECCs, this scheme also has an obvious disadvantage, that is, the noise
parameters $\lambda ^{\left( i\right) }$ in the Hamiltonian (1) should be
known accurately and must not change in an unknown way.

A crucial assumption for this scheme is that two qubits can be set close so
that they are decohered collectively. Ref. [36] shows this is the case if
distance $d$ between the two qubits satisfies $d<<\overline{\lambda }$,
where $\overline{\lambda }$ is the mean effective wave length of the noise
field. In practice, such as in the ion-trapped quantum computers, where the
noise is from the thermal variation of the qubits [23], this assumption
seems reasonable. It is now well understood that quantum errors are harder
to correct than classical errors, since there appear new kinds of errors,
such as the phase errors and the bit-phase errors. Here we show, if we have
some knowledge of the interaction of the qubits with the environment,
quantum errors are easier to prevent. This supports a commonplace, but
fundamentally important, observation that the more one knows about the
noise, the easier it is to correct for it.\\

{\bf Acknowledgment}

This project was supported by the National Natural Science Foundation of
China.

\newpage\

\end{document}